\documentclass[aps
floats,prd,nofootinbib,superscriptaddress,10pt]{revtex4-1}
\usepackage[utf8]{inputenc}
\usepackage{bm}
\usepackage{amsmath}
\usepackage{comment}
\usepackage{caption}
\usepackage{color}

\usepackage{graphicx,amsmath,amsfonts,amssymb
}

\usepackage[english]{babel}

\begin{document}

\title{Constraints on small-scale primordial power\\ by annihilation signals from extragalactic dark matter minihalos}
\author{Tomohiro Nakama}

\affiliation{Department of Physics and Astronomy, Johns Hopkins
     University, 3400 N.\ Charles St., Baltimore, MD 21218, USA}

\author{Teruaki Suyama}

\affiliation{Research Center for the Early Universe (RESCEU),
Graduate School of Science, The University of Tokyo, Tokyo 113-0033, Japan}

\author{Kazunori Kohri}

\affiliation{
Institute of Particle and Nuclear Studies, KEK, 1-1 Oho, Tsukuba, Ibaraki 305-0801, Japan}

\affiliation{The Graduate University for Advanced Studies (SOKENDAI), 1-1, Oho, Tsukuba, Ibaraki
305-0801, Japan}

\affiliation{Rudolf Peierls Centre for Theoretical Physics, The University of
Oxford, 1 Keble Road, Oxford OX1 3NP, UK}

\author{Nagisa Hiroshima}

\affiliation{
Institute of Particle and Nuclear Studies, KEK, 1-1 Oho, Tsukuba, Ibaraki 305-0801, Japan}

\affiliation{
Institute for Cosmic Ray Research, The University of Tokyo, 5-1-5 Kashiwanoha, Kashiwa,
Chiba 277-8582, Japan}

\begin{abstract}
We revisit constraints on small-scale primordial power from annihilation signals from dark matter minihalos. Using gamma rays and neutrinos from extragalactic minihalos and assuming the delta-function primordial spectrum, we show the dependence of the constraints on annihilation modes, the mass of dark matter, and the annihilation cross section. We report both conservative constraints by assuming minihalos are fully destructed when becoming part of halos originating from the standard almost-scale invariant primordial spectrum, and optimistic constraints by neglecting destruction. 
\end{abstract}

\maketitle

\section{Introduction}
Primordial fluctuations of a vast range of wavelengths are expected to have been generated in the early universe. While their power spectrum on large scales has been well determined \cite{Ade:2015xua}, that on small scales has been constrained only weakly by the null detection of primordial black holes \cite{Carr:1994ar,Kim:1996hr,Green:1997sz,Zaballa:2006kh,Carr:2009jm,Nakama:2013ica,Nakama:2014fra} (see also Refs. \cite{Kohri:2007qn,Peiris:2008be} and references therein), induced gravitational waves \cite{Saito:2008jc,Saito:2009jt,Assadullahi:2009jc,Alabidi:2012ex,Alabidi:2013wtp,Nakama:2016gzw}, the spectral distortions of the cosmic microwave background \cite{Chluba:2012we,Nakama:2016kfq,Nakama:2017ohe}, and acoustic reheating \cite{Jeong:2014gna,Nakama:2014vla}. 

Since the abundance of dark matter minihalos is also sensitive to the small-scale primordial power, microlensing \cite{Ricotti:2009bs} as well as astrometric lensing \cite{Li:2012qha} and Shapiro time delay \cite{Clark:2015sha,Clark:2015tha}
they could cause can also be used to constrain small-scale power. See also  Refs. \cite{Kohri:2014lza,Aslanyan:2015hmi}. In addition, since the density of dark matter is relatively high inside minihalos, annihilation there could be substantial if dark matter particles annihilate. Hence, observations of gamma rays \cite{Josan:2010vn,Bringmann:2011ut} or neutrinos \cite{Yang:2013dsa} can also be used to constrain small-scale primordial power, which depends on the properties of dark matter such as their mass, annihilation cross section and annihilation modes. 

In Ref. \cite{Bringmann:2011ut}, upper limits on primordial power are obtained using only minihalos formed before some collapse redshift, denoted by $z_c$ and chosen by hand, and hence their limits depend on $z_c$ (tighter limits for smaller $z_c$). This method has also been adapted in subsequent papers. 
In principle, we should be able to derive upper limits on primordial power without introducing such additional parameter $z_c$, 
which we attempt in this paper focusing on gamma rays and neutrinos from minihalos. 

We consider some spike in the primordial spectrum superposed on the standard almost-scale-invariant spectrum, 
and constrain its amplitude by annihilation signals from minihalos. The spike is assumed to be sufficiently substantial such that halos originating from the feature form at relatively high redshifts, and later, halos originating from the standard spectrum also form. For convenience, in this paper we call the former class of halos ultracompact minihalos (UCMHs) 
and the latter standard halos. 

At low redshifts when standard structures form, some UCMHs are accreted onto larger halos that form in the standard $\Lambda$CDM scenario.
Simple calculation shows that UCMHs trapped in the larger halo collide with each other many times in the Hubble time.
In this process, they are likely to loose a substantial fraction of their masses \cite{Giocoli:2008qg}, 
and annihilation signals from them would somewhat diminish. 
In \cite{Bringmann:2011ut}, this effect was neglected, arguing that the density is high enough in the central cores of UCMHs, yielding dominant contributions to annihilation signals, for those cores to survive under their assumptions.
They adopted a halo profile of $r^{-9/4}$, predicted by the simple secondary infall model where spherical symmetry is assumed \cite{Bertschinger:1985pd}, which was argued to hold well for halos formed at high redshifts. 
This model and setting $z_c$ sufficiently high would justify the neglect of the aforementioned potential destruction effects. 
However, according to recent simulations \cite{Gosenca:2017ybi, Delos:2017thv},
this simple expectation does not hold and inner density profile of minihalos is much shallower. 

In this paper, we aim to derive upper limits on small-scale power without introducing $z_c$, 
which implies we also use minihalos formed at relatively low redshifts. 
In order to give conservative bound on the primordial power, instead of using the steep profile $\propto r^{-9/4}$,
we adopt the Navarro–Frenk–White (NFW) profile \cite{Navarro:1995iw} for minihalos, confirmed for standard halos 
relatively well by N-body simulations. 
Using minihalos formed at lower redshifts assuming the NFW profile, 
the internal density of UCMHs is low relative to UCMHs formed at high redshifts considered in the literature, 
and as a result the destruction effects would be more important. 
Since to what extent destruction affects the annihilation rate inside UCMHs is difficult to quantify precisely, 
we focus on annihilation signals from extragalactic minihalos, 
and adopt both the most conservative assumption that minihalos are fully destructed when accreted onto larger, 
standard halos, and also the optimistic assumption that the destruction effects are fully negligible, 
thereby obtaining conservative and optimistic upper limits on primordial power on small scales. 
Furthermore, we extend the previous studies by studying how the upper limit on the
primordial power changes as we change the unknown factors such as annihilation cross section, 
dark matter mass, and annihilation channels.
We consider both gamma rays and neutrinos to derive the constraints.

\section{Methodology}
Primordial density perturbations with large amplitude on small scales produce UCMHs.
Once formed, pair-annihilations of dark matter inside UCMHs continuously take place and the high energy cosmic rays
are emitted out of them.
In this section, we briefly review the methodology of how to connect the cosmic-ray flux with the
primordial density perturbations.

\subsection{Mass function of UCMHs}
The first thing to do is to compute the mass function of UCMHs from the given primordial density perturbations.
We define $dn/dM (M,z)dM$ by the proper number density of UCMHs in the mass interval $(M,M+dM)$.
Thus, its normalization is given by
\begin{equation}
\int M \frac{dn}{dM}dM=\frac{3H_0^2}{8\pi G} \Omega_{m0} {(1+z)}^3.
\end{equation}

We use the Press-Schechter formalism \cite{Press:1973iz} to obtain the mass function of UCMHs 
originating from some spike of primordial power on small scales as follows. 
Let us introduce the smoothed dark matter density contrast $\delta_M({\vec x})$:
\begin{equation}
\delta_M({\vec x}) = \int d^3{\vec x'} ~\delta ({\vec x'})
W(|{\vec x}-{\vec x}'|;R),
\end{equation}
where $\delta$ is the density contrast evaluated in the linear perturbation
theory, $W(r;R)$ is the window function (both $r$ and $R$ are comoving distances) 
and $M$ is the total mass of the dark matter enclosed in an effective volume
specified by the window function.
In this paper, we use the Gaussian window function \cite{Zentner:2006vw}
\begin{equation}
W(r;R)={(2\pi)}^{-3/2}R^{-3} \exp \left( -\frac{r^2}{2R^2} \right),\\
\end{equation}
and its Fourier transform is 
${\tilde W} (k;R) \equiv\int d^3x~e^{-i {\vec k} \cdot {\vec x}} W({\vec x};R)=e^{-k^2R^2/2}$.
The window volume is defined as the inverse of the window function evaluated at $r=0$, and
for the Gaussian window function, it is $V_{W}={(2\pi)}^{3/2} R^3.$
Defining the mass of a collapsed object corresponding to a smoothing scale $R$ as
\begin{equation}
M(R) = \int d^3 x ~V_W W(r;R)~\rho_{m0} = \rho_{m0} V_W,
\end{equation}
the mass for the Gaussian window function is given by
\begin{equation}
M={(2\pi)}^{3/2} \rho_{m0}R^3.~~~~~
\end{equation}

Given any $\delta_M$ at redshift $z$,
regions where $\delta_M({\vec x},z)$, evaluated in the linear theory, exceeds a critical value $\delta_c$ 
can roughly be regarded as part of collapsed objects more massive than $M$, according to the spherical collapse model \cite{Weinberg:2008zzc}.
The value of $\delta_c$ is about $1.686$ for the Einstein-de Sitter universe
and is insensitive to the presence of dark energy \cite{Eke:1996ds}.
Thus, we use $\delta_c=1.686$ throughout this paper.

Assuming Gaussianity of the density contrast,
the total fraction of such regions is given by
\begin{equation}
\beta(M,z)=\frac{1}{\sqrt{2\pi}\sigma_M(z)}\int_{\delta_c}^\infty d\delta \exp\left(-\frac{\delta^2}{2\sigma_M^2(z)}\right)
=\frac{1}{\sqrt{2\pi}}\int_\nu^\infty dx\exp\left(-\frac{x^2}{2}\right)=\frac{1}{2}\mathrm{erfc}\left(\frac{\nu}{\sqrt{2}}\right),
\end{equation}
where $\sigma_M (z)$ is the standard deviation of $\delta_M (z)$
and is related to the linear matter power spectrum $P_{\delta} (k,z)$ by
\begin{equation}
\sigma_M^2(z)=\int \frac{dk}{2\pi^2} k^2 P_{\delta}(k,z) {\tilde W}^2 (k;R),
\end{equation}
and $\nu_M(z)\equiv\delta_c/\sigma_M(z)$. 
The matter perturbation $\delta$ is related to the primordial potential $\Phi_p$ by \cite{Dodelson:2003ft}
\begin{equation}
\delta(\bm{k},a)=\frac{3}{5}D_1(a)\frac{k^2T(k)}{\Omega_mH_0^2}\Phi_p(\bm{k}).
\end{equation}
Here, $D_1$ is the growth function given by \cite{Belloso:2011ms}
\begin{equation}
D_1(a)=a\cdot {}_2F_1\left[\frac{w-1}{2w},\frac{-1}{3w},1-\frac{5}{6w};1-\Omega_m^{-1}(a)\right],
\end{equation}
where $w$ is a constant dark energy equation of state parameter, here set to $w=-1$,
${}_2F_1$ is the Hypergeometric function, and $\Omega_m^{-1}(a)=1+\Omega_\Lambda a^3/\Omega_{m0}$.
In addition, $T(k)$ is the transfer function, and one may use the zero-baryon transfer function of 
Ref. \cite{Eisenstein:1997ik}, 
\begin{equation}
T(q)=\frac{\ln (2e+1.8q)}{\ln (2e+1.8q) +\left( 14.2+\frac{731}{1+62.5q} \right)q^2},~~~~~~
q=\frac{k}{\Omega_{m0}h^2 {\rm Mpc}^{-1}},
\end{equation}
which was noted to be more accurate and simpler than the so-called BBKS transfer function of Ref.\cite{Bardeen:1985tr}. 
Notice that this transfer function ignores both free streaming and kinetic decoupling of dark matter
and becomes invalid above a cutoff scale $k_c$ where those effects are important.
Essentially, dark matter halos are not formed out of perturbations with $k>k_c$.
For typical WIMP dark matter with canonical cross section $\langle \sigma v \rangle =3\times 10^{-26}{\rm cm^3/s}$, 
$k_c \sim 10^{6\sim 7} {\rm Mpc}^{-1}$ \cite{Green:2005fa}.
It is possible that the cross section in the present universe is different from the one in the early universe \cite{Griest:1990kh}.
In the next section, we study the dependence of the constraint on the cross section and 
this cross section refers to the value in the present universe.
The primordial potential $\Phi_p$ is related to the comoving curvature perturbation ${\cal R}$ via ${\cal R}=3\Phi_p/2$ \cite{Dodelson:2003ft}. Hence (see also Ref. \cite{Baghram:2014nha}),
\begin{equation}
{\cal P}_\delta(k,a)=\frac{4}{25}\left[D_1(a)\frac{k^2T(k)}{\Omega_{m0}H_0^2}\right]^2{\cal P}_{{\cal R}}(k).
\end{equation}

The Press-Schechter mass function is obtained by differentiating $\beta$ with respect to the mass and multiplying 
by $\rho_m/M$ with a fudge factor $2$;
\begin{equation}
\frac{dn}{d\ln M}=-2\frac{\rho_m}{M}\frac{d\beta}{d\ln M}=\sqrt{\frac{2}{\pi}}\frac{\rho_m}{M}\frac{d\nu_M}{d\ln M}\exp\left(-\frac{\nu_M^2}{2}\right)
=\sqrt{\frac{2}{\pi}}\frac{\rho_m}{M}\frac{d\ln\sigma_M^{-1}}{d\ln M}\nu_M\exp\left(-\frac{\nu_M^2}{2}\right).
\end{equation}
Integration over $M$ yields a relation,
\begin{equation}
\int_0^\infty dM\frac{dn}{d\ln M}=2\rho_m\beta(M=0,z), \label{nint}
\end{equation}
which will be used later.

\begin{figure}[tp]
\begin{center}
\includegraphics[width=10cm,keepaspectratio,clip]{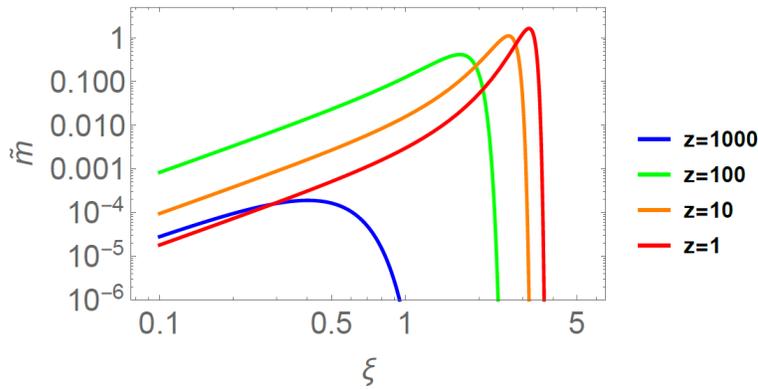}
\end{center}
\caption{The dimensionless mass function $\tilde{m}$ defined by Eq.~(\ref{mtilde}) for
$k_*=10^5{\rm Mpc}^{-1}, {\cal A}=10^{-3}$ at four different redshifts $z=10^3, 100, 10, 1$.
}
\label{mass-function}
\end{figure}

In this paper, we consider the following delta-function type spectrum:
\begin{equation}
{\cal P}_{{\cal R}} (k)={\cal A}^2k \delta (k-k_*).
\end{equation}
For this power spectrum, we find 
\begin{equation}
\sigma_{M}(z)=\sigma_0(z)\tilde{W}(k_*;R),
\quad\sigma_0(z)\equiv\frac{2{\cal A}D_1(z)}{5\Omega_{m0}H_0^2}k_*^2T(k_*)\label{sigma0}.
\end{equation}
Introducing $\xi\equiv k_*R$ and $\tilde{\delta}\equiv \delta_c/\sigma_0$, 
the mass function can be rewritten as
\begin{align}
\tilde{m}\equiv\frac{M}{\rho_{m0}}\frac{dn}{d\ln M}&=\sqrt{\frac{2}{\pi}}\frac{d\xi}{d(\ln \xi^{3})}\frac{d}{d\xi}\left(\frac{\xi^2}{2}\right)\tilde{\delta}e^{\xi^2/2}\exp \left(-\frac{\tilde{\delta}^2}{2}e^{\xi^2}\right)\nonumber\\
&=\frac{1}{3}\sqrt{\frac{2}{\pi}}\tilde{\delta}\xi^2\exp\left(\frac{\xi^2}{2}-\frac{\tilde{\delta}^2}{2}e^{\xi^2}\right).\label{mtilde}
\end{align}
This non-dimensional mass function $\tilde{m}$ represents the number of halos in the volume $M/\rho_{m0}$ per logarithmic interval in $M$, that is, 
if it is order unity for some $\xi$, most of the dark matter is 
part of halos whose mass corresponds to that $\xi$.
Fig.\ref{mass-function} shows $\tilde{m}$ for
$k_*=10^5{\rm Mpc}^{-1}, {\cal A}=10^{-3}$ as a function of $\xi$ at four different redshifts $z=10^3, 100, 10, 1$.
We find that halos are rare at $z=10^3$ but have been produced abundantly by $z \sim 100$
with their size distribution sharply peaked at around $\xi \simeq 1$.
Typical size of halos $\xi$ continues to increase slightly subsequently.

We use the NFW profile $\rho_h=\rho_s/[(r/r_s)(1+r/r_s)^2]$, which has two parameters.
Let us define the radius of halos $R_h$ as the virial radius.
Then, according to the spherical collapse model, the virial radius is given by
\begin{equation}
R_h= {\left( \frac{3M}{4\pi \Delta \rho_m} \right)}^{1/3}
\end{equation}
where $\rho_m$ is the cosmic mean dark matter density and $\Delta$ is given by \cite{Bryan:1997dn}:
\begin{equation}
\Delta
\simeq (18\pi^2+82x-39 x^2)/\Omega_m(t),
\end{equation}
where $x=\Omega_m(t)-1$.
The free parameters $\rho_s$ and $r_s$ appearing in the NFW profile can be fixed as follows.
From the volume integration of the NFW profile up to the virial radius, we find that $\rho_s$
can be written as
\begin{equation}
\rho_s=\frac{c^3}{\ln (1+c)-\frac{c}{1+c}} \frac{\Delta \rho_m}{3}.
\end{equation}
Fixing $r_s$ is more non-trivial.
Numerical simulations suggest that the concentration parameter $c$ defined by
$c\equiv R_h/r_s$ evolves as a function of the cosmic time as \cite{Zhao:2008wd}
\begin{equation}
c=4\times\left[1+\left(\frac{t}{3.75t_{0.04}}\right)^{8.4}\right]^{1/8},
\end{equation}
where $t_{0.04}$ is the time when a UCMH gained $4\%$ of its mass evaluated at time $t$.
We adopt the above fitting formula for the concentration parameter.
As Fig.\ref{mass-function} shows, for a delta-function primordial spectrum, $\tilde{m}$ is sharply peaked, 
and therefore we assume that, at each moment, UCMHs have the same concentration, 
and that $t_{0.04}=t_{0.04}(t)$ is roughly given by the time when the peak mass was the $4\%$ of the peak mass at the time $t$.
Under this simplifying assumption, $t_{0.04}$ is determined uniquely by $t$, $k_*$ and ${\cal A}$.
Let us explain in more detail how this program works.
To this end, we denote the position of the peak of $\tilde{m}$ by $\xi_m$. 
Then $\xi_m$ and $\tilde{\delta}$ are related via
\begin{equation}
\frac{d(\ln\tilde{m})}{d\xi}\bigg|_{\xi_m}=\frac{2}{\xi_m}+\xi_m-\tilde{\delta}^2\xi_m e^{\xi_m^2}=0,
\quad \tilde{\delta}=\left[\frac{e^{-\xi_m^2}}{\xi_m}\left(\frac{2}{\xi_m}+\xi_m\right)\right]^{1/2}\equiv g(\xi_m).
\end{equation}
For a given $\tilde{\delta}$, $\xi_m(z)$ can be determined by solving the above equation. Then the redshift $z_{0.04}$, when the peak mass was $4\%$ of the peak mass at $z$, is obtained by solving $\xi_m[z_{0.04}(z)]=0.04^{1/3}\xi_m(z)$, and then $t_{0.04}(z)=t[z_{0.04}(z)]$.
The time $t$ and the redshift are related via 
\begin{equation}
H_0t=\int_0^a\frac{da}{a}\left(\Omega_\Lambda+\frac{\Omega_{m0}}{a^3}
\right)^{-1/2}=
\frac{2}{3}\Omega_\Lambda^{-1/2}\mathrm{arcsinh}\left[\left(\frac{\Omega_\Lambda}{\Omega_{m0}}a^3\right)^{1/2}\right].
\end{equation}

\subsection{Cosmic rays from UCMHs}
The intensity of gamma-rays or neutrinos from extragalactic UCMHs may be written as \cite{Campbell:2011kf,Allahverdi:2011sx}
\begin{equation}
I(E)=\langle \sigma v \rangle \int\frac{dz}{H(z)}W[(1+z)E,z]\langle\rho^2\rangle(z).
\end{equation}
Here, 
\begin{equation}
W(E,z)=\frac{1}{8\pi m_\chi^2}\frac{1}{(1+z)^3}\frac{dN}{dE}(E)e^{-\tau(E,z)},
\end{equation}
where $m_\chi$ is the mass of dark matter, $dN/dE$
is the energy spectrum of particles arising from one annihilation, calculated using PYTHIA 8.2 \cite{Sjostrand:2006za,Sjostrand:2007gs,Sjostrand:2014zea}, and the factor $e^{-\tau}$, for the case of gamma rays,  takes into account absorption due to the extragalactic background light and the cosmic microwave background. 
We use the "fiducial" model of Ref. \cite{Gilmore:2011ks} for absorption due to the extragalactic background light and 
use Ref. \cite{Stecker:2005qs} for absorption due to the cosmic microwave background (see also Ref. \cite{Stecker}). Note that, in these references, $\tau=\tau(E_0,z)$ is provided in terms of the observed energy $E_0$, whereas $E$ in eq. (2) is the energy in the rest frame of the sources. For the case of neutrinos, we neglect attenuation i.e. $e^{-\tau}\simeq 1$ \cite{Murase:2012xs}. 
In addition,
\begin{equation}
\langle\rho^2\rangle(z)=\int dM\frac{dn}{dM}(M,z)\int d^3r\rho_h^2(r|M,z),
\end{equation}
where the volume integral is performed inside the radius of halos $R_h$.
Using Eqs. (\ref{nint}) and an equation (see also Refs. \cite{Taylor:2002zd,Zavala:2009zr}
)
\begin{equation}
4\pi\int_0^{R_h}\rho_h^2r^2dr=\Delta\rho_mMf(c),\quad
f(c)\equiv
\frac{c^3[1-1/(1+c^3)]}{9[\ln(1+c)-c/(1+c)]^2}, \label{rhosqint}
\end{equation}
we find
\begin{equation}
\langle\rho^2\rangle(z)=2\Delta\beta(M=0) f(c)\rho_m^2.
\end{equation}

\section{constraints on primordial small-scale power from minihalos}

As stated in the Introduction, we constrain primordial power by conservatively assuming UCMHs are fully destructed 
when they become part of halos that form later in the standard $\Lambda$CDM scenario, 
and also by optimistically assuming destruction is fully negligible. 
To this end, let us introduce the field fraction $\beta_f$, the fraction of UCMHs which are field halos or which have not become subhalos of standard halos, and multiply $\langle\rho^2\rangle$ above by this factor, when we adopt the conservative assumption of full destruction. The field fraction is given by $\beta_f(M,z)=1-2\beta_{\mathrm{st}}(M,z)$, where $2\beta_{\mathrm{st}}$ is the fraction of the matter inside collapsed objects larger than mass $M$ at redshift $z$, and is calculated from 
the following standard primordial spectrum: 
\begin{equation}
{\cal P}_{{\cal R}}={\cal A}^2_{\mathrm{st}}\left(\frac{k}{k_p}\right)^{n_s-1},
\end{equation}
where ${\cal A}^2_{\mathrm{st}}=2.2\times 10^{-9}, \,n_s=0.97$ and $k_p=0.05\,\mathrm{Mpc}^{-1}$ \cite{Ade:2015xua}. The field fraction depends on the mass of UCMHs, but for $\beta_f$ we simply set the mass to the peak mass of the mass function $\tilde{m}$ of UCMHs, which is determined by $z,\,{\cal A}$ and $k_*$.

The gamma-ray intensity from UCMHs is compared with the isotropic diffuse gamma-ray background, and we obtain upper limits on ${\cal A}$, as a function of $k_*$ so that the former does not exceed the latter in the energy range in which the latter is inferred. 
In Ref. \cite{Ackermann:2014usa} they parameterize the isotropic diffuse gamma-ray background in an energy range between 100 MeV and 820 GeV as
\begin{equation}
\frac{dN}{dE}=I_{100}\left(\frac{E}{100\,\mathrm{MeV}}\right)^{-\gamma}\exp\left(\frac{-E}{E_{\mathrm{cut}}}\right),
\end{equation}
and for their foreground model "A" they found
$I_{100}=(0.95\pm 0.08)\times 10^{-7}\mathrm{MeV}^{-1}\mathrm{cm}^{-2}\mathrm{s}^{-1}\mathrm{sr}^{-1}$, $\gamma=2.32\pm 0.02$, $E_{\mathrm{cut}}=279\pm 52 \,\mathrm{GeV}$. We conservatively set $I_{100}=1.11\times 10^{-7}\mathrm{MeV}^{-1}\mathrm{cm}^{-2}\mathrm{s}^{-1}\mathrm{sr}^{-1}$, $\gamma=2.28$, $E_{\mathrm{cut}}=383\,\mathrm{GeV}$. 

\begin{figure}[htp]
\begin{center}
\includegraphics[width=13cm,keepaspectratio,clip]{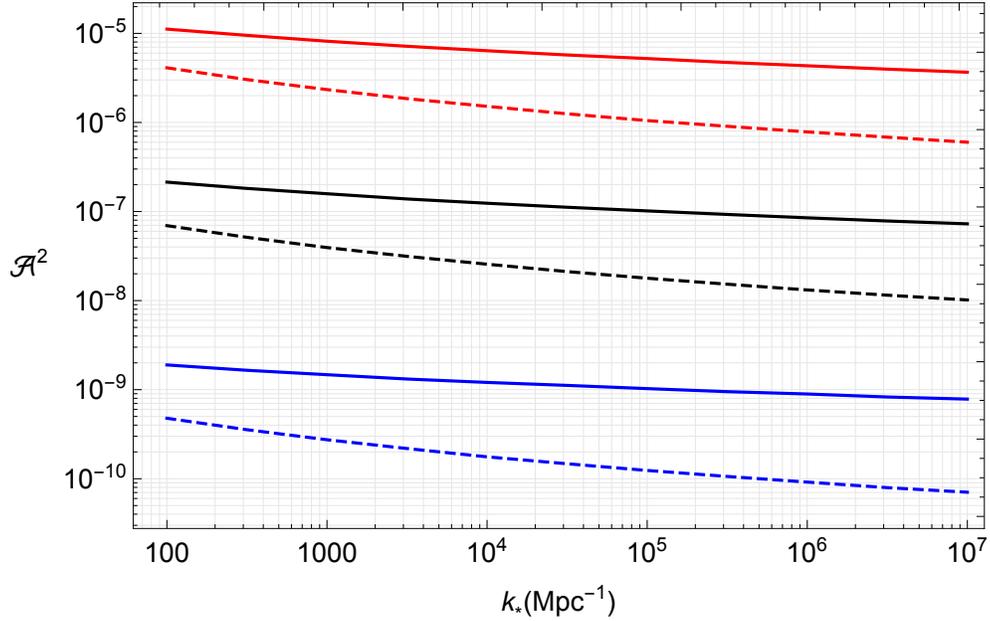}
\end{center}
\caption{
Constraints from gamma rays for the $b\bar{b}$ mode and $m_\chi=1\mathrm{TeV}$, with (solid) and without (dashed) the field fraction multiplied. From top to bottom, $\langle \sigma v \rangle
=\{10^{-3},1,10^3\}\langle\sigma v\rangle_{\mathrm{can}},
$ where $\langle\sigma v\rangle_{\mathrm{can}}=3\times 10^{-26}\mathrm{cm}^3\mathrm{s}^{-1}$.
}
\label{mail1}
\end{figure}

\begin{figure}[htp]
\begin{center}
\includegraphics[width=13cm,keepaspectratio,clip]{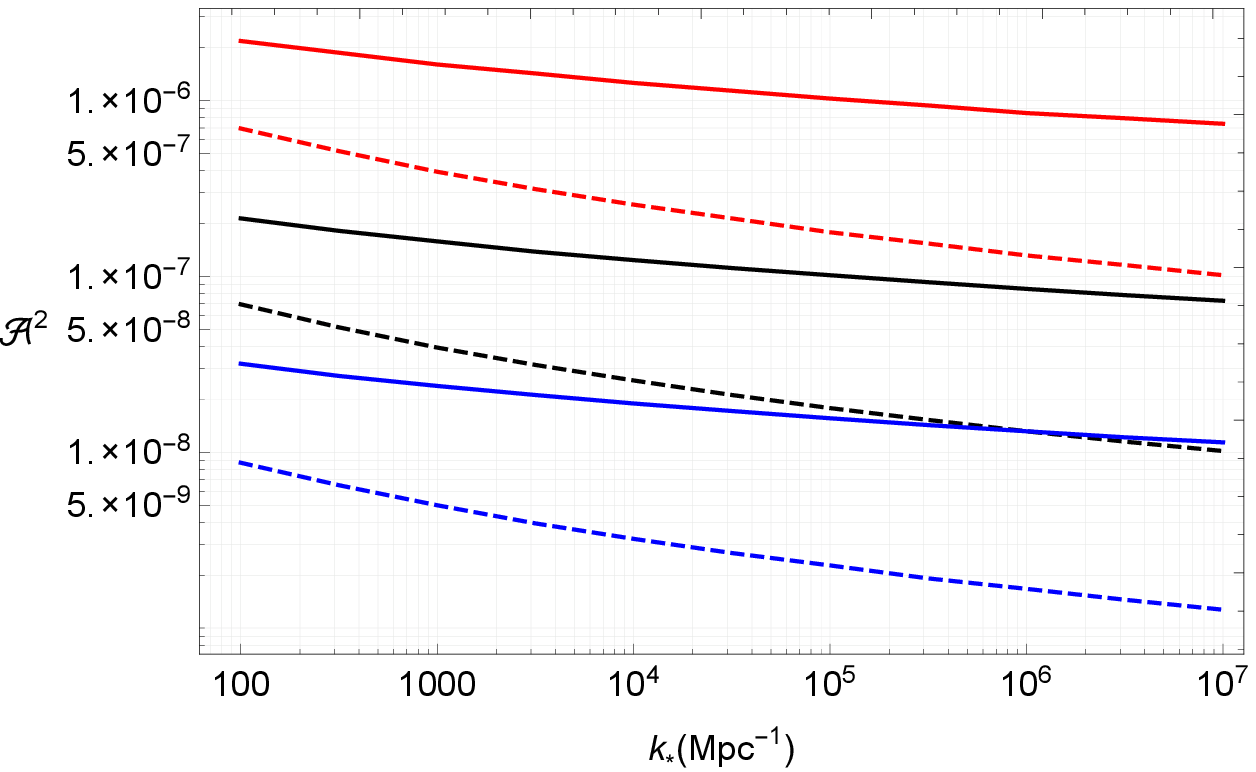}
\end{center}
\caption{
Constraints from gamma rays, for the $b\bar{b}$ mode and $\langle\sigma v\rangle =3\times 10^{-26}\mathrm{cm}^3\mathrm{s}^{-1}$. The dark matter mass $m_\chi$ is 0.01 TeV (blue), 1 TeV (black) and 100 TeV (red). 
}
\label{mail2}
\end{figure}

\begin{figure}[htp]
\begin{center}
\includegraphics[width=13cm,keepaspectratio,clip]{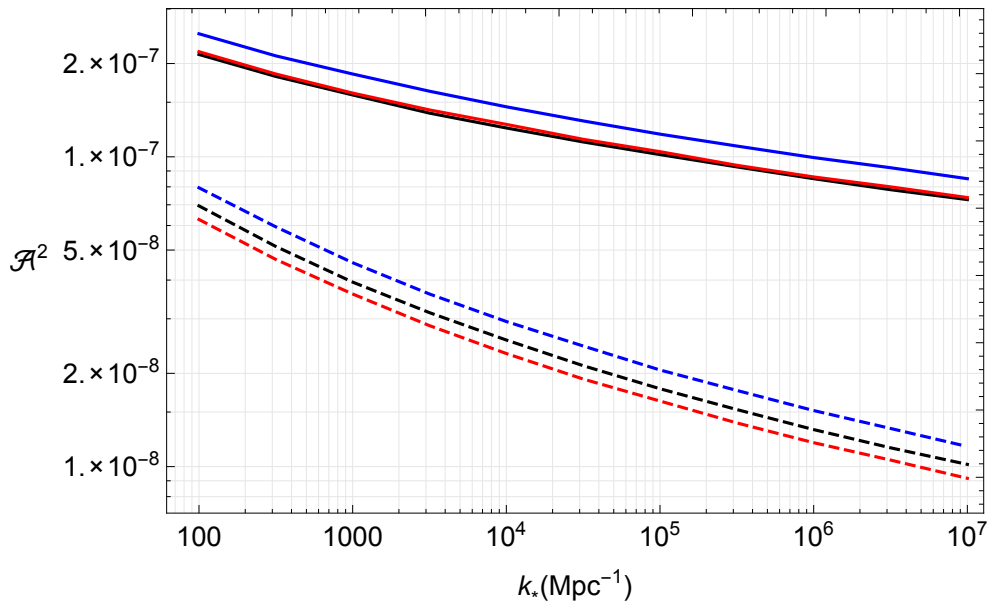}
\end{center}
\caption{
Constraints from gamma rays for $m_\chi=1$ TeV and $\langle \sigma v \rangle = 3\times 10^{-26}\mathrm{cm}^3\mathrm{s}^{-1}$. The annihilation mode is $b\bar{b}$ (black), $W^+W^-$ (blue) and $\tau^+\tau^-$ (red). 
}
\label{mail3}
\end{figure}

The upper limit on ${\cal A}$ as a function of $k_*$ obtained from the observations of gamma rays
are shown in Fig. \ref{mail1} (for three different values of $\langle \sigma v \rangle$),
Fig. \ref{mail2} (for three different values of dark matter mass $m_\chi$), 
and Fig. \ref{mail3} (for three different annihilation channels). 
From Fig. \ref{mail1}, we find that when $\langle\sigma v\rangle$ is larger, 
only a smaller number of UCMHs is allowed, hence tighter constraints (Fig. \ref{mail1}),
which is a reasonable result. 
From Fig. \ref{mail2}, we find that the constraints become tighter for smaller $m_\chi$. 
This can be understood as the result of the boosted the annihilation rate due to
larger number density of dark matter particles.  
In Fig. \ref{mail3}, constraints are plotted for three annihilation channels
$\to b {\bar b}$, $\to W^+ W^-$, and $\to \tau^+ \tau^-$, assuming that the annihilations
are dominated by either one of the channels.
It is clear from this figure that the constraints depend only weakly on the annihilation channels we consider.

\begin{figure}[htp]
\begin{center}
\includegraphics[width=13cm,keepaspectratio,clip]{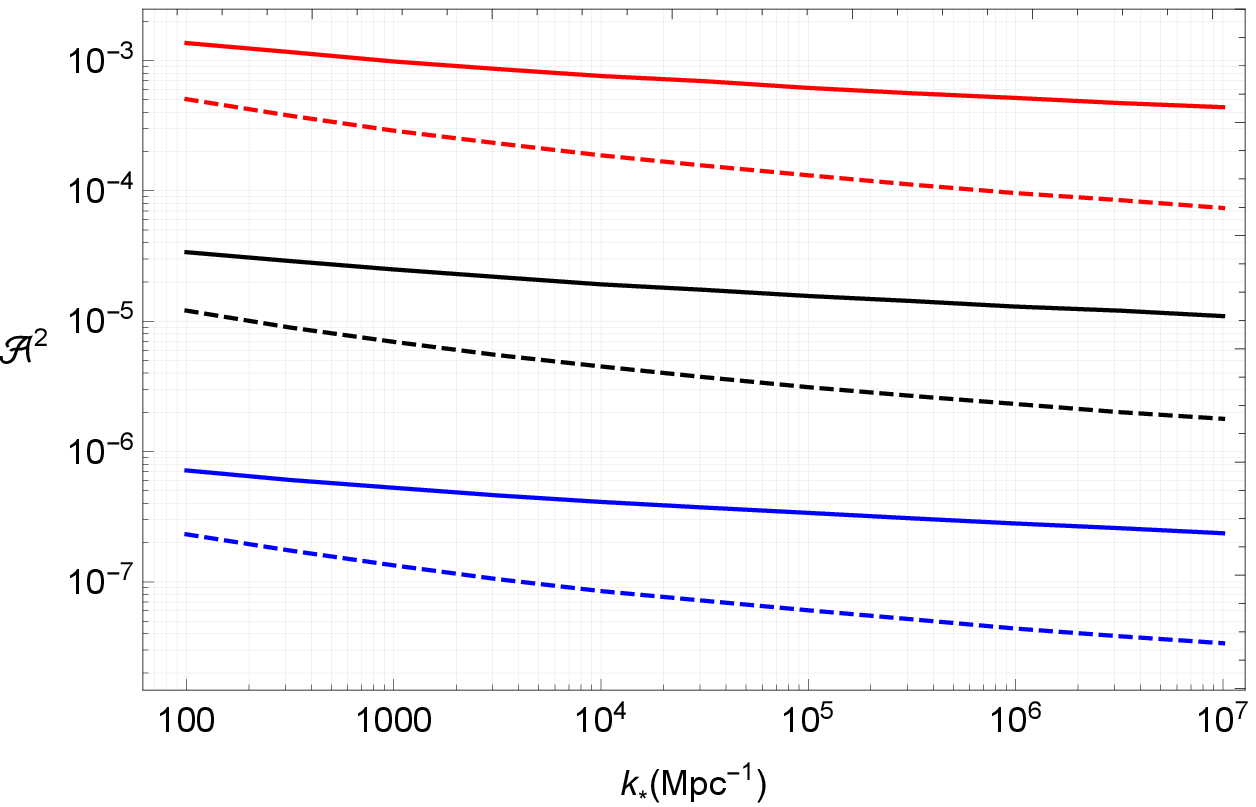}
\end{center}
\caption{
Constraints from neutrinos for the $b\bar{b}$ mode and $m_\chi=1\mathrm{TeV}$, with (solid) and without (dashed) the field fraction multiplied. From top to bottom, $\langle \sigma v \rangle
=\{10^{-3},1,10^3\}\langle\sigma v\rangle_{\mathrm{can}},
$ where $\langle\sigma v\rangle_{\mathrm{can}}=3\times 10^{-26}\mathrm{cm}^3\mathrm{s}^{-1}$.
}
\label{mail4}
\end{figure}

\begin{figure}[htp]
\begin{center}
\includegraphics[width=13cm,keepaspectratio,clip]{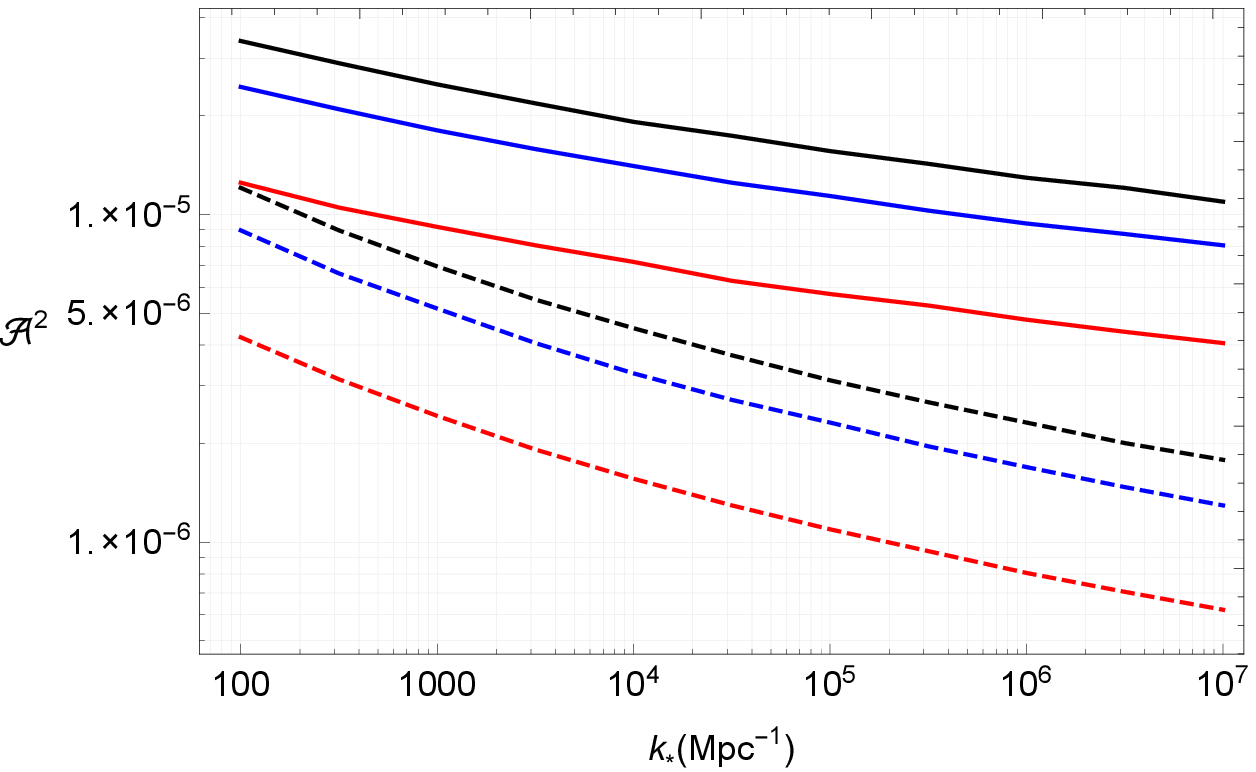}
\end{center}
\caption{
Constraints from neutrinos, for the $b\bar{b}$ mode and $\langle\sigma v\rangle =3\times 10^{-26}\mathrm{cm}^3\mathrm{s}^{-1}$. The dark matter mass $m_\chi$ is 0.01 TeV (blue), 1 TeV (black) and 100 TeV (red). 
}
\label{mail5}
\end{figure}

\begin{figure}[htp]
\begin{center}
\includegraphics[width=13cm,keepaspectratio,clip]{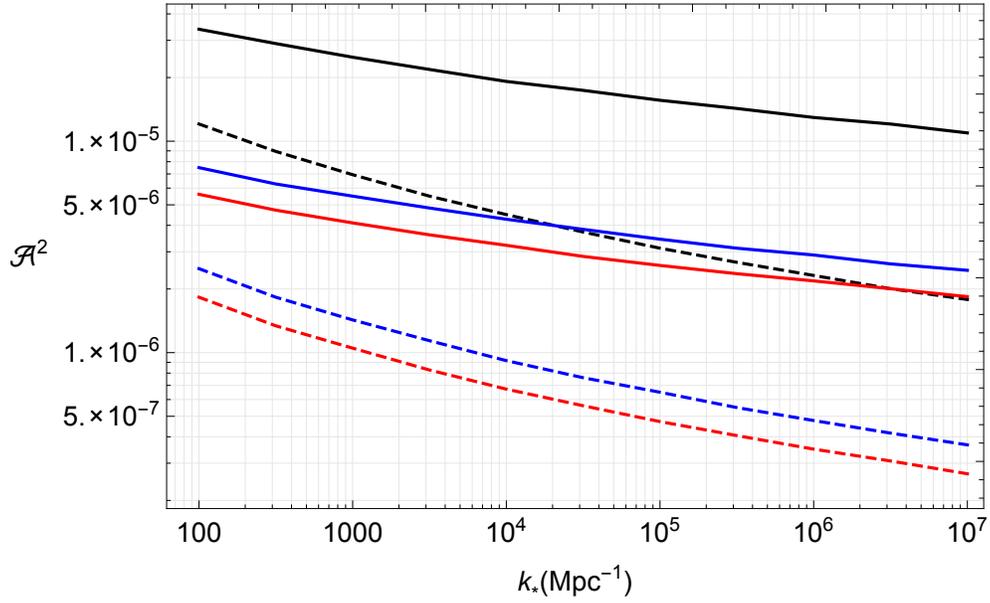}
\end{center}
\caption{
Constraints from neutrinos for $m_\chi=1$ TeV and $\langle \sigma v \rangle = 3\times 10^{-26}\mathrm{cm}^3\mathrm{s}^{-1}$. The annihilation mode is $b\bar{b}$ (black), $W^+W^-$ (blue) and $\tau^+\tau^-$ (red). 
}
\label{mail6}
\end{figure}

For the case of neutrinos, 
we refer to Ref. \cite{Lee:2012pz} to take into account neutrino oscillation: 
$\Phi_{\nu_\mu}=0.24\Phi_{\nu_e}^0+0.40\Phi^0_{\nu_\mu}+0.35\Phi^0_{\nu_\tau}$, 
where $\Phi^0_{\nu_i}$ denotes neutrino fluxes at the sources. 
We obtain upper limits on ${\cal A}^2$ as a function of $k_*$, 
requiring that the muon neutrino intensity from UCMHs be less than the intensity of atmospheric muon neutrinos. 
We use the angle-dependent atmospheric muon neutrino intensity shown in Ref. \cite{Erkoca:2010vk}, setting the zenith angle to zero. 
The constraints from neutrinos are shown in Figs. \ref{mail4} - \ref{mail6} in exactly the same manner
as in the case of gamma rays. 
From these figures, we find that in all cases the constraints from neutrinos are weaker than those from gamma rays.
While Figs.\ref{mail1} and \ref{mail4} look similar, Fig. \ref{mail5} qualitatively differs from Fig. \ref{mail2},
namely, the constraints become stronger as the dark matter mass is increased from $1 {\rm TeV}$
to $100 {\rm TeV}$.
Although the dark matter number density becomes smaller as we increase the dark matter mass, 
this also shifts the energy spectrum toward higher energies where the atmospheric neutrino intensity is smaller. 
These two effects determine the dependence of constraints on $m_\chi$ (Fig. \ref{mail5}).



\section{conclusion}
We have revisited constraints on small-scale primordial power from annihilation signals from dark matter minihalos. Using gamma rays and neutrinos from extragalactic minihalos and assuming the delta-function primordial spectrum, we show the dependence of the constraints on annihilation modes, the mass of dark matter, and the annihilation cross section. We report both conservative constraints by assuming minihalos are fully destructed when becoming part of halos originating from the standard almost-scale invariant primordial spectrum, and optimistic constraints by neglecting destruction. 

In the literature, they have introduced some collapse redshift $z_c$, taking into account only contributions from UCMHs formed before $z_c$, to derive constraints on primordial power, which then depends on the choice of $z_c$. Here we have attempted to derive constraints without introducing $z_c$, which implies we also use UCMHs formed at lower redshifts. Hence, we used the NFW profile instead of the profile predicted by the simple secondary infall model ($\rho\propto r^{-9/4}$), adopted in the literature of UCMHs and argued to hold well for UCMHs formed at sufficiently high redshifts. 

Our constraints obtained using gamma rays from UCMHs depend on the properties of extragalactic background light, have not been fully understood. As we gain more understanding about them through future observations or simulations, this uncertainty would hopefully diminish. Constraints from neutrinos are weaker but cleaner in this regard.

\section{acknowledgement}
This work is supported in
part by Grant-in-Aid for JSPS Fellow No. 25.8199 and JSPS Postdoctoral Fellowships
for Research Abroad (T.N.), MEXT KAKENHI Nos.~JP15H05889, and JP16H00877 (K.K.),  
JSPS KAKENHI Nos. 26247042 and JP1701131 (K.K.), 
JP17H06359 (T.S.), JP15H05888 (T.S.), and JP15K17632 (T.S.).

\end{document}